\documentclass[namedreferences]{solarphysics}
\usepackage[hyperref, optionalrh, natbib]{spr-sola-addons} 
\usepackage{graphicx}                    
\usepackage{color}   
\usepackage{breakurl}                    

\newcommand{\degree}{^{\circ}}

\newcommand{\Rsun}{R$_{\odot}$}
 
\newcommand{\eg}{\textit{e.g.~}}
\newcommand{\ie}{\textit{i.e.~}}
\newcommand{\insitu}{{\it in situ}}

\begin{document}

\begin{article}

\begin{opening}

\title{Coronal Mass Ejections from the Same Active Region Cluster: Two Different Perspectives}

%
\author{H.~\surname{Cremades}$^{1}$\sep
        C.H.~\surname{Mandrini}$^{2,3}$\sep
        B.~\surname{Schmieder}$^{4}$\sep
        A.M.~\surname{Crescitelli}$^{5}$
       }

%
\runningauthor{Cremades et al.}
\runningtitle{CMEs from the Same Active Region Cluster}

%
\institute{$^{1}$ Universidad Tecnol\'ogica Nacional - Facultad Regional Mendoza and CONICET, Mendoza, Argentina\\
                  email: \url{hebe.cremades@frm.utn.edu.ar}\\ 
		     $^{2}$ Instituto de Astronom\'ia y F\'isica del Espacio (IAFE), CONICET-UBA, Buenos Aires, Argentina\\                    
           $^{3}$ Facultad de Ciencias Exactas y Naturales (FCEN), UBA, Buenos Aires, Argentina\\
                  email: \url{mandrini@iafe.uba.ar} \\
           $^{4}$ Observatoire de Paris, Meudon 92190, France\\
           		  email: \url{brigitte.schmieder@obspm.fr}\\
           $^{5}$ Instituto Balseiro - Universidad Nacional de Cuyo, San Carlos de Bariloche, Argentina\\
           		  email: \url{alberto.crescitelli@ib.edu.ar}
          }

\begin{abstract}
The cluster formed by active regions (ARs) NOAA 11121 and 11123, approximately located on the solar central meridian on 11 November 2010, is of great scientific interest. This complex was the site of violent flux emergence and the source of a series of Earth-directed events on the same day. The onset of the events was nearly simultaneously observed by the 
{\it Atmospheric Imaging Assembly} (AIA) telescope aboard the {\it Solar Dynamics Observatory} (SDO) and the 
{\it Extreme-Ultraviolet Imagers} (EUVI) on the {\it Sun-Earth Connection Coronal and Heliospheric Investigation} (SECCHI) suite of telescopes onboard the {\it Solar-Terrestrial Relations Observatory} (STEREO) twin spacecraft. The progression of these events in the low corona was tracked by the {\it Large Angle Spectroscopic Coronagraphs} (LASCO) onboard the {\it Solar and Heliospheric Observatory} (SOHO) and the SECCHI/COR coronagraphs on STEREO. ̣̣SDO and SOHO imagers provided data from the Earth's perspective, whilst the STEREO twin instruments procured images from the orthogonal directions. This spatial configuration of spacecraft allowed optimum simultaneous observations of the AR cluster and the coronal mass ejections that originated in it. Quadrature coronal observations provided by STEREO revealed a notably large amount of ejective events compared to those detected from Earth's perspective. Furthermore, joint observations by SDO/AIA and STEREO/SECCHI EUVI of the source region indicate that all events classified by GOES as X-ray flares had an ejective coronal counterpart in quadrature observations.
These results have direct impact on current space weather forecasting because of the probable missing alarms when there is a lack of solar observations in a view direction perpendicular to the Sun-Earth line.
\end{abstract}
%
\keywords{Coronal mass ejections, initiation and propagation; Coronal mass ejections, low coronal signatures; Prominences, dynamics}
\end{opening}


%
\section{Introduction}
\label{s:Introduction} 

Long before coronal mass ejections (CMEs) were discovered as outward-traveling density enhancements in the field of view of coronagraphs four decades ago (\opencite{Tousey-etal1974, Gosling-etal1974}), their existence was proposed to explain alterations in the Earth's magnetosphere (\opencite{Carrington1859, Lindemann1919, Alfven1956, Morrison1956}). Later this association was confirmed and CMEs were established as key space weather modulators (\eg \opencite{Gosling1993}). In spite of this substantiation and after few decades of space weather research, the prediction of the occurrence of a CME eruption from a specific solar region is still not possible. Therefore, the only plausible way to perform CME space-weather forecasting until now is based on the case of an Earth-directed CME and, additionally, on related phenomena like low-coronal signatures and radio waves (\eg \opencite{Pick-etal2006}). At present, such a forecasting requires {\it sine qua non} the detection of a CME that is at least partially traveling in the Earth's direction. 

	CMEs are three-dimensional entities (\eg \opencite{Howard-etal1982, Crifo-etal1983, Webb1988}), with some of them proven to be organized along an axial direction, holding cylindrical symmetry (\opencite{Cremades-Bothmer2004, Moran-Davila2004}). This configuration agrees with the existence of helical magnetic fields coiled in a cylindrical, croissant-like shape, evidenced as circular threads outlining the dark cavity in several CMEs with favourable orientation (\eg \opencite{Dere-etal1999, Gibson-Low2000, Vourlidas-etal2013}). It is plausible that all CMEs contain these magnetic flux ropes (\opencite{Vourlidas-etal2013}) capable of interacting with Earth's magnetosphere depending on the orientation of their magnetic field, although this constitutes a fundamental matter of debate. As CMEs are three-dimensional optically thin structures captured in two-dimensional images, their appearance in the field of view of a coronagraph depends on the viewing angle. Observations of CMEs from multiple viewpoints, available for the past seven years, allow coarse three-dimensional reconstruction of these entities through a variety of techniques that have different limitations (see the reviews by \opencite{Mierla-etal2010} and \opencite{Thernisien-etal2011}).

When CMEs travel along the Sun-Earth line, they appear as a halo surrounding the coronagraph's occulter (\opencite{Howard-etal1982}), whose symmetry depends on: i) how close the CME propagation direction lies with respect to the Sun-Earth line and ii) the shape of the CME's outer envelope. Only some few halo CME events per year appear symmetric, bright, and having an uninterrupted circular or elliptical leading edge at all position angles (see \citet{Lara-etal2006} for a discussion on the nature of these particular events), while most halo CMEs often appear as ragged, diffuse, and faint structures. Because of the Thomson scattering effect (\eg \opencite{vandeHulst1950, Hundhausen1993, Andrews2002, Vourlidas-Howard2006, Howard-DeForest2012}), plasma material is seen to shine brighter when its location is such that the radius from the Sun is normal to the line of sight. Points fulfilling this condition shape the Thomson sphere. Close to the Sun, this can be approximated to material lying close to the plane of the sky (POS), \ie the plane perpendicular to the line of sight. Therefore, if an observer detecting a faint and ragged halo CME could change vantage point 90$\degree$ away, he would see the CME traveling in his respective POS, significantly brighter and with an essentially different shape (\eg like a lightbulb). So far, there is no appraisal of the amount of CMEs that turn out to be so faint when Earth-directed that they are undetectable by current coronagraphs. As mentioned above, CMEs are generally considered potentially geoeffective if they are Earth-directed, \ie exhibiting a halo or a partial halo in coronagraph images. It is unfortunate that these events are poorly detected and their attributes hard to be inferred from Earth's perspective. Although there is a wealth of techniques to correct for projection effects in kinematic and morphological parameters from Earth's perspective (\eg \opencite{Zhao-etal2002, Michalek-etal2003, Xie-etal2004, Cremades-Bothmer2005, Vrsnak-etal2007, HowardT-etal2008, Temmer-etal2009}), all of them have important inherent uncertainties and thus hinder space weather tasks. 

Solar sources of Earth-directed CMEs offer unbeatable views due to their location close to central meridian (CM), but without observations offset with respect to the Sun-Earth line, most of the times it is hard or even impossible to claim the source of a CME eruption only based on low-coronal or chromospheric observations. In this article, we investigate the punctual case of the complex formed by active regions (ARs) NOAA 11121 and 11123, located close to CM on 11 November 2010, and the CMEs that arose from it. The CM location makes the cluster a perfect object of study, so that a wealth of space- and Earth-borne solar instrumentation was able to register its phenomena in a variety of wavelengths and regimes. On that day, the occurrence of a number of eruptions could be deduced from low-coronal data. Based on coronagraph data from Earth's perspective, only one halo CME was reported on that day. However, images of the solar corona provided by the {\it Solar-Terrestrial Relations Observatory}  (STEREO) spacecraft, located almost 180$\degree$ appart on that date, reveal a surprisingly large amount of Earth-directed CME eruptions. Joint observations of the source region procured by the {\it Solar Dynamics Observatory} (SDO) and STEREO were key to find the precise location and time of each eruption in the chromosphere and low corona, and at the same time unexpectedly disclosed that all events classified by GOES as X-ray flares had an ejective coronal counterpart in quadrature observations. This study highlights the importance of solar monitoring space missions that are located offset from the Sun-Earth line, \eg at the L4 or L5 Lagrange points, not only for the practical purpose of space weather predictions, but also to deepen the understanding of the physical processes involved in CME eruptions.

\section{Instrumentation and Data}\label{s:Data} 

Data captured by spacecraft on a quadrature configuration for the date of interest, covering a variety of regimes, are crucial for this study. Images from Earth's perspective were recorded by the {\it Solar Dynamics Observatory} (SDO; \opencite{Pesnell-etal2012}) and by the {\it Solar and Heliospheric Observatory} (SOHO; \opencite{Domingo-etal1995}). The quadrature views were provided by instruments on the {\it Solar-Terrestrial Relations Observatory} (STEREO; \opencite{Kaiser-etal2008}). The STEREO mission consists of two identical spacecraft following Earth's orbit around the Sun, one ahead of it (ST-A) and the other behind it (ST-B). The spacecraft separate at a rate of $\approx$ 45$\degree$ per year, which implies that for 11 November 2010 the separation between them was $\approx$180$\degree$.

The Earth-facing solar source at low-coronal and photospheric levels was investigated through images provided by SDO's {\it Atmospheric Imaging Assembly} (AIA; \opencite{Lemen-etal2012}) and {\it Helioseismic and Magnetic Imager} (HMI; \opencite{Schou-etal2012}) experiments, respectively. Its chromospheric aspect was analyzed aided by H$\alpha$ images from the Paris-Meudon spectroheliograph (\url{http://bass2000.obspm.fr/}). Orthogonal views of the AR cluster, \ie at the solar limb, were provided by the {\it Extreme-Ultraviolet Imagers} (EUVI) on the {\it Sun-Earth Connection Coronal and Heliospheric Investigation} (SECCHI; \opencite{Howard-etal2008}) onboard STEREO.  CMEs in the white-light corona as viewed from Earth were identified in data provided by SOHO's {\it Large Angle Spectroscopic Coronagraph} (LASCO; \opencite{Brueckner-etal1995}) C2 and C3 coronagraphs, which cover the ranges from 1.7 to 6.0 and from 3.7 to 32 solar radii, respectively. Quadrature observations of the solar white-light corona were provided by the COR1 and COR2 coronagraphs of the STEREO/SECCHI investigations suite, which cover 1.4 to 4.0 and 2.0 to 15 solar radii of the solar corona.

A number of CME catalogs and data bases were consulted as an alternative way to obtain the number of CMEs arising from the complex of ARs facing Earth. These are the SOHO/LASCO CME Catalog (\url{http://cdaw.gsfc.nasa.gov/CME_list}) (\opencite{Yashiro-etal2004}), the Computer Aided CME Tracking catalog (\url{http://sidc.oma.be/cactus/}) (CACTus; \opencite{Robbrecht-Berghmans2004}), the Solar Eruptive Event Detection System (\url{http://spaceweather.gmu.edu/seeds/}) (SEEDS; \opencite{Olmedo-etal2008}), the Coronal Image Processing (\url{http://alshamess.ifa.hawaii.edu/CORIMP/}) (CORIMP; \opencite{Byrne-etal2012}), and the Automatic Recognition of Transient Events and Marseille Inventory from Synoptic maps (\url{http://cesam.lam.fr/lascomission/ARTEMIS/}) (ARTEMIS; \opencite{Boursier-etal2009}). Each of these automated catalogs rely on different techniques to recognize CME events, which determine their respective limitations.

\section{Events on 11 November 2010}
\subsection{The Cluster of Active Regions}
\label{GOES-description}

\begin{figure} 
\includegraphics[width=0.99\textwidth]{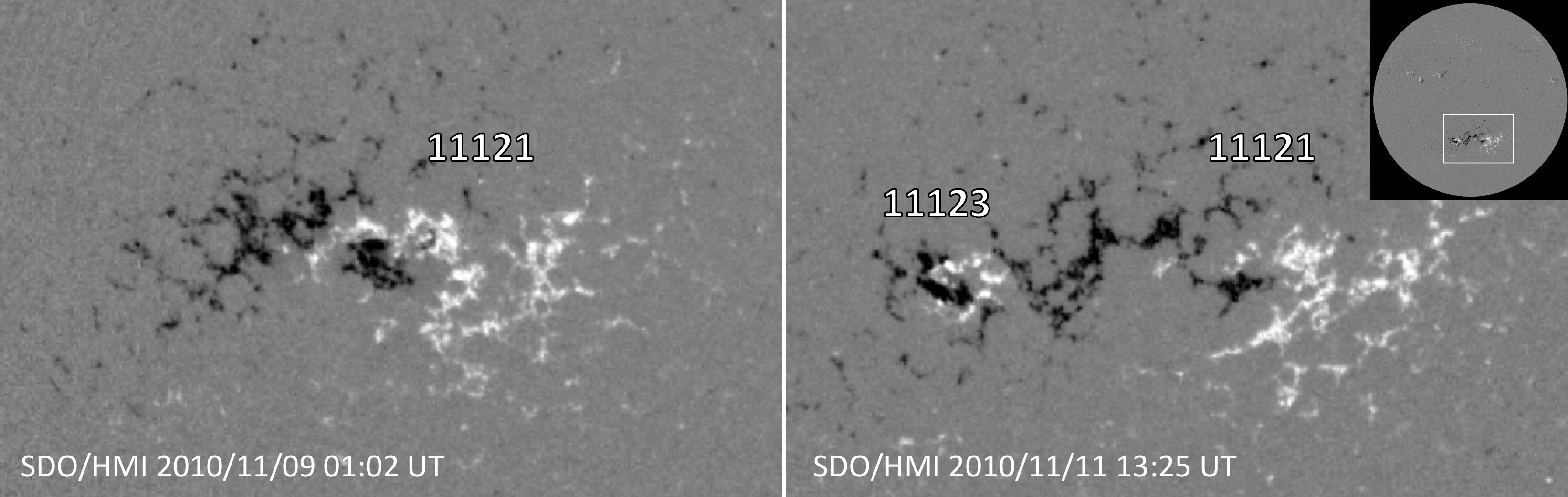}
\caption{SDO/HMI magnetograms: AR 11121 on 9 November 2010, before the emergence of AR 11123 (left panel), and the AR cluster on 11 November 2010, after the emergence of AR 11123 (right panel).}
\label{fig:HMI}
\end{figure}

From 9 to 11 November 2010, close to CM and at $\approx$ 21$\degree$ in the southern solar hemisphere, AR 11123 rapidly emerged within the following negative polarity of AR 11121 that had persisted for five solar rotations \citep{Mandrini-etal2014}. In just one day the total magnetic flux increased by 70\% with the emergence of the new bipole groups associated to AR 11123. 
Figure~\ref{fig:HMI} shows the region of interest before and after the emergence in images of the photospheric magnetic field recorded by the SDO/HMI magnetograph. Both ARs formed a complex, consisting of several nested bipoles by 11 November. On this date, a wealth of solar X-ray flares were recorded by the {\it Geostationary Operational Environmental Satellites} (GOES). Several of them were C-class flares, mostly occurring within AR 11123 and some also in conection with AR 11121. Motivated by this sequence of events, \inlinecite{Mandrini-etal2014} throughly analyzed this cluster of regions by implementing a magnetic field topology analysis that let them explain the locations of flare ribbons for several events. 

\begin{figure} 
\centerline{\includegraphics[width=0.9\textwidth,clip=]{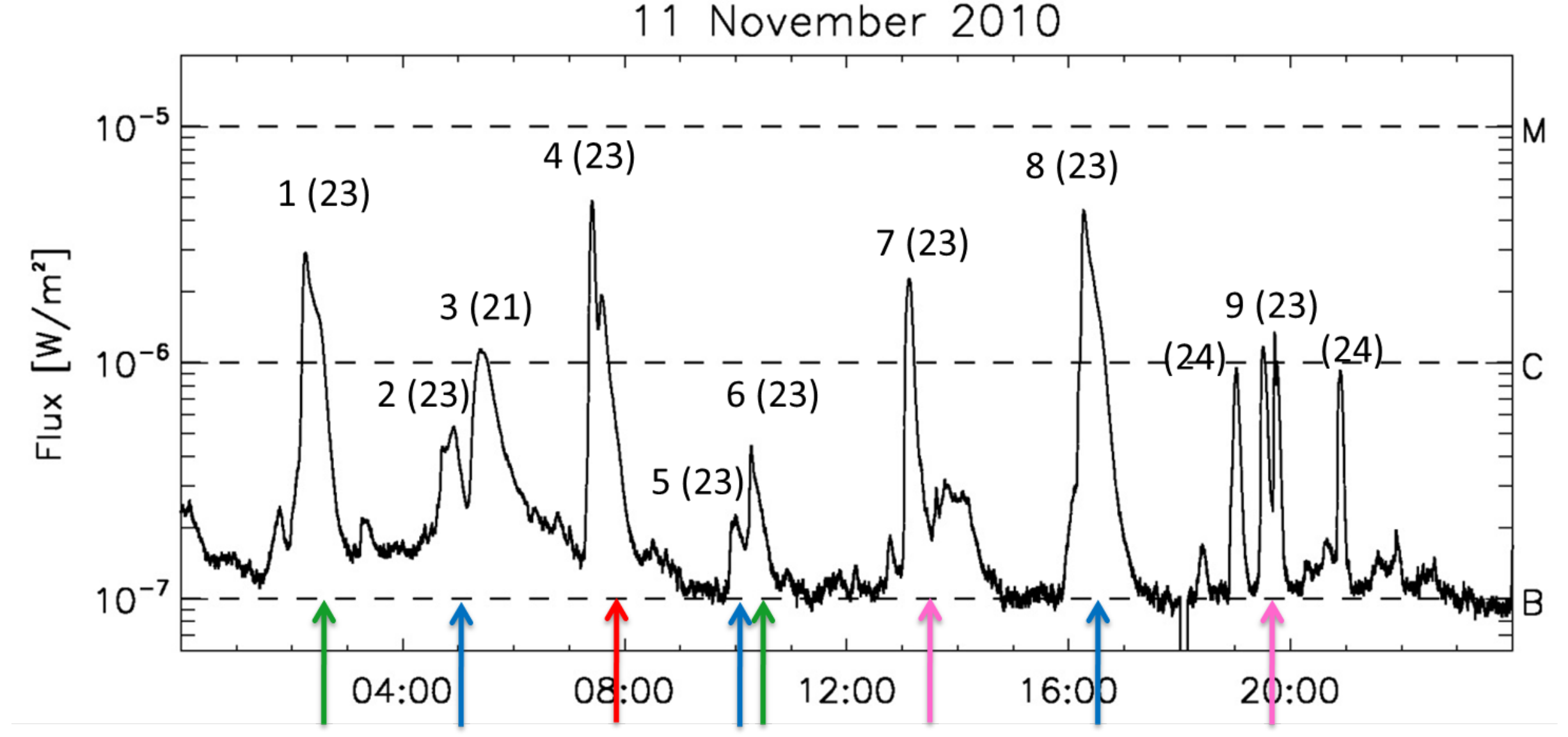}}
\caption{GOES X-ray flux for 11 November 2010. Numbers close to the peaks identified in GOES records indicate the flare number as in Table~\ref{tbl:flares}, while those in brackets represent the last two digits of the AR number where the flare took place. Arrow colours indicate homologous flares.}
\label{fig:GOES}
\end{figure}

\begin{figure} 
\centerline{\includegraphics[width=0.9\textwidth,clip=]{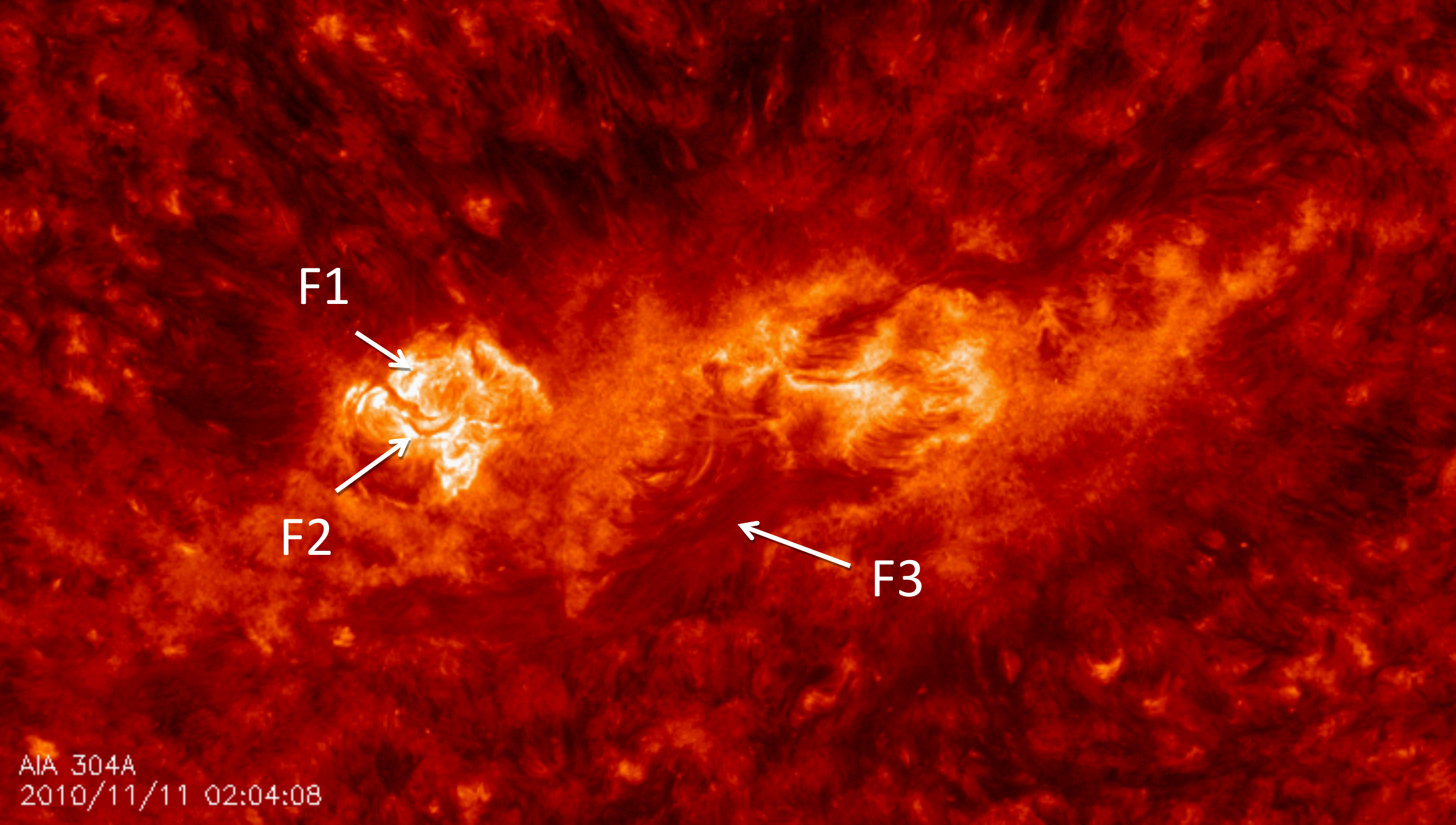}}
\caption{AIA 304\AA~snapshot of the AR cluster. The image shows the detectable filaments involved in some of the eruptions of 11 November 2010.}
\label{fig:filaments}
\end{figure}

Figure~\ref{fig:GOES} and Table~\ref{tbl:flares} summarize all X-ray flare activity identified in GOES records and related to the AR cluster on 11 November 2010. Some events consist of more than one intense peak (see, \eg, FL4 and FL9 in Figure~\ref{fig:GOES}) which were not classified separately in GOES records; for these cases, GOES classification refers to the most intense X-ray peak. Homologous flares, \ie~flares that are morphologically very similar, are indicated in the figure by arrows of the same colour. The first column of Table~\ref{tbl:flares} assigns an identificative number to each GOES flare, which will be used later for the associations with the observed CMEs. Columns 2 to 4 represent the flare most intense peak time, location, and X-ray class, respectively. The last column refers to comments on the flare and related filament eruption activity. Most of the flares were C-class and appeared limited to AR 11123, although loops interconnecting both of the ARs composing the complex were seen to expand in connection to eruptive activity. Three of the GOES flares in Table~\ref{tbl:flares} (FL3, FL4, and FL8) were associated to conspicuous eruptions involving filaments, labeled in Figure~\ref{fig:filaments}. An in-depth analysis of the double peak exhibited by FL4 in GOES curve can be found in \inlinecite{Mandrini-etal2014}. These authors modeled in detail the magnetic field topology of the AR complex. They identified two main magnetic null-points (see their Figure 9). Magnetic reconnection at one of the null points, and its associated separatrices and quasi-separatrix layers (QSLs; \opencite{Demoulin-etal1996}), was proposed as the origin of the pre-flare brightenings observed before the first GOES peak. This peak corresponds to a clearly visible eruptive event. The second GOES peak, with maximum at 07:35 UT, corresponds to a confined event. These authors suggested that this event was probably initiated by reconnection at the second null point; this process was  forced by the destabilization of a portion of F1 (Figure~\ref{fig:filaments}) that was not observed to erupt during the previous GOES peak. Futhermore, the topology explaining the pre-flare brightness observed before FL8 is also shown in Figure 15 of \inlinecite{Mandrini-etal2014}, while the flare itself has been analyzed in detail by \inlinecite{Huang2014}. Flares FL1, FL2, FL7, and FL9 are related to small eruptions, detected in high-cadence running difference images of SDO/AIA 171\AA, 193\AA, and 304\AA. No evidence of eruption was found in connection with flares FL5 and FL6, while no eruptive signatures were observed away from the flare time-windows at all. 

\begin{table}
\caption{GOES X-ray flare activity related to the AR cluster on 11 November 2010. Column one assigns a number to each flare (FL) as in Figure~\ref{fig:GOES}. Columns two to four correspond to the reported GOES flare peak-time and X-ray class, and location, respectively. Column five includes comments on the flare and related filament-eruption activity.}
\label{tbl:flares}
\begin{tabular}{llllp{6cm}}     
\hline
FL & \multicolumn{3}{c}{GOES X-ray flares} & Comments and related activity\\
\hline
1 & 02:14 & C2.9 & S24E16 & Two-ribbon flare, small eruption to the SE of AR 11123, no filament seen.\\
2 & 04:55 & B5.4 &  & Two-ribbon flare at both sides of filament F1, faint eruption to the NE of AR 11123.\\ 
3 & 05:24 & C1.1 &  & North portion of filament F3 seen when erupting from AR 11121.\\
4 & 07:25 & C4.7 & S27E14 & First peak (07:25 UT), two-ribbon flare, eruption of a portion filament F2 to the south of AR 11123. Second peak (07:35 UT), confined event (see text).\\
5 & 10:00 & B2.2 &  &  Brightening to the NE of AR 11123, no evidence of eruption.\\
6 & 10:16 & B4.3 &  & Brightening to the SE of AR 11123, no evidence of eruption.\\
7 & 13:07 & C2.2 &  & Two-ribbon flare, eruption of faint filament reformed in the site of F2.\\
8 & 16:16 & C4.3 & S26E08 & Two-ribbon flare, eruption of filament F1, eruption of a portion of quiescent filament F3 to the south of the AR complex.\\
9 & 19:30 & C1.1 & S26E06 & Two-ribbon flare, eruption of faint filament reformed in the site of F2.\\
\hline
\end{tabular}
\end{table}

The activity and location of the complex of ARs implies an increased potential of production of Earth-directed events that motivates a close inspection of the ensuing ejective events. The results of investigating coronagraph data recorded from the Earth's perspective, which is the case of the SOHO/LASCO experiment, are presented in the next subsection. The differing outcome from the inspection of coronagraph data registered 90$\degree$ away from the Sun-Earth line, \ie by the STEREO/SECCHI instrument suite, is presented subsequently.

\subsection{SOHO/LASCO CME Detections on 11 November 2010}\label{s:SOHOCMEs} 

After a number of flares accompanied by filament eruptions were found to happen in low-coronal images of the AR cluster, and given the fact that the AR cluster is located at central meridian, it would be expected that some halo/partial-halo CMEs show up in images of the corona captured by the SOHO/LASCO coronagraph, situated in the Sun-Earth line. As it was anticipated in Section~\ref{s:Data}, five CME catalogs were consulted in order to assess their performance in the detection of events that have an important component of their propagation direction toward Earth. Table~\ref{tbl:LASCOCMEs} summarizes the detections reported by these catalogs on 11 November 2010. CME events occurring in the north--west corona have not been listed in the table for simplicity reasons, given that the AR complex is located in the southern hemisphere. The first column of Table~\ref{tbl:LASCOCMEs} designates a CME number, while the following columns indicate the time (UT) at which each catalog first detected each CME. An empty cell means that CME was not detected by that catalog, while two entries on the same cell mean that the same CME was detected as two different events by a catalog. The number in brackets is the event's angular width (in degrees) as listed in each catalog. The systematically smaller values of angular width listed in automated catalogs can be explained by limitations of automatic recognition algorithms. CACTus does not appear in the table because it did not detect any CME on this date. The SOHO/LASCO catalog reports five distinct CMEs, all of them also detected by SEEDS. In addition, SEEDS reports two more events, apparently considered in the SOHO/LASCO catalog as trailing material of previous events. Events at 20:24 UT and 21:48 UT from SEEDS may refer to the same  SOHO/LASCO CME at 20:12 UT. For its part, CORIMP reports three CMEs on this date, but two of them are part of the same SOHO/LASCO and SEEDS CME at 17:00 UT. ARTEMIS reports three of the CMEs detected by SOHO/LASCO and SEEDS, with start time systematically later than the one informed by these catalogs.

\begin{table}
\caption{CME detections by SOHO-based catalogs related to the AR cluster on 11 November 2010. SO as first column label refers to CMEs detected by coronagraphs onboard SOHO. The first column assigns a number to each CME, while the following columns indicate the time (UT) at which each catalog first detected each CME. The number in brackets is the angular width (in degrees) of each CME as measured by the different catalogs. See Section~\ref{s:SOHOCMEs} for further explanations.}
\label{tbl:LASCOCMEs}
\begin{tabular}{lllll}     
\hline
SO & SOHO/LASCO & SEEDS & CORIMP & ARTEMIS \\
\hline
1 & 08:24 (74) & 08:48 (31) & 08:36 (12) & 10:18 (12) \\
2 & 11:00 (33) & 11:36 (12) & & 12:43 (14) \\
3 & 14:00 (111) & 14:48 (17) & & \\
4 & 17:00 (260) & 17:00 (80) & 17:00 (67) & 17:32 (60) \\
&&& 17:00 (7) &\\
5 & & 19:00 (9) & & \\
6 & 20:12 (38) & 20:24 (15)& & \\
&& 21:48 (10) & & \\
7 & & 22:12 (11) & & \\
\hline
\end{tabular}
\end{table}

In summary, only two CMEs were reported by all catalogs, while only one event is reported as a partial-halo CME by the SOHO/LASCO catalog. This outcome is far from expected, given the number of eruptions observed in low coronal data, and the CM location of the AR complex. Except for the one partial-halo event, all the other CMEs arising from the AR complex were faint, ragged, narrow, and asymmetric. In a first attempt, events FL4, FL7, FL8, and FL9 of Table~\ref{tbl:flares} can be temporally associated to CMEs SO1, SO3, SO4, and SO6 of Table~\ref{tbl:LASCOCMEs}. Low-coronal eruptive events FL1, FL2, and FL3 of Table~\ref{tbl:flares} would have no observable coronal counterparts, while some of the reported CMEs in Table~\ref{tbl:LASCOCMEs}, if not an artifact from automated catalogs, would not have been expected from the observed low coronal activity. However, the deduction of associations only based on temporal agreements is a tricky task. The position angle of the CME, \ie its angular central location measured counterclockwise from the solar north, should also be in accordance with the solar source location; this is a difficult association to do if the source region is close to CM. Therefore, in order to clarify the disagreements found between Tables~\ref{tbl:flares} and~\ref{tbl:LASCOCMEs}, data from the twin STEREO spacecraft are inspected. The STEREOs vantage points, 90$\degree$ appart from the Sun-Earth line, allow the observation of the AR complex on their respective solar limbs, and therefore have the best view of the CMEs that originated there, since they travel on their respective POS.

\subsection{STEREO/SECCHI CME Detections on 11 November 2010}\label{s:STEREOCMEs} 

Examination of SECCHI data was originally meant to aid in establishing associations between the flares attributed to the AR cluster and the CMEs detected from Earth's perspective. However, SECCHI data was revealing not only in regard to this aspect, but also to the recognition of deficiencies in the monitoring of Earth-directed events. Several ejective events not evident from Earth's perspective were discerned; these were not hinted by the ejective activity observed in SDO/AIA. The particular configuration of the STEREO spacecraft allowed us to observe the AR cluster on their respective solar limbs, \ie on the west limb of ST-B and on the east limb of ST-A. Events listed in Table~\ref{tbl:SECCHICMEs} are unambiguously distinguished and observed to be born in the AR cluster. These are seen travelling outward in the COR1 and COR2 coronagraphs, both A and B, with an important propagation component toward Earth. Column one in Table~\ref{tbl:SECCHICMEs} allots a number for each event, while columns two and three indicate the start time of the CME in the COR1-A and COR2-A instruments, respectively. The number in brackets next to the start time in column two represents the angular width in degrees of each CME, as measured in COR1-A images. Given the particular configuration of the STEREO spacecraft with respect to the cluster of ARs, the angular width of the ensuing CMEs is nearly the same from the perspective of the ST-A and ST-B coronagraphs considering that these CMEs travel in their POS. Column four indicates the flare number assigned in Table~\ref{tbl:flares} to each GOES flare related to the AR cluster, whereas column five refers to the CME number given in Table~\ref{tbl:LASCOCMEs}. The last column includes comments on the sources of eruptions and on related phenomena. 

\begin{table}
\caption{CMEs born in the AR cluster on 11 November 2010 detected with STEREO 
instruments. ST as first column label refers to CMEs detected by coronagraphs onboard STEREO. 
Column one gives a number to each event. Columns two and three indicate the start time of the 
CME in the COR1-A and COR2-A instruments, respectively. The number in brackets next to the 
start time in column two is the angular width in degrees of each CME measured in COR1-A images. 
Column four indicates the flare number in Table~\ref{tbl:flares}, while column five refers 
to the CME number given in Table~\ref{tbl:LASCOCMEs}.}
\label{tbl:SECCHICMEs}
\begin{tabular}{lllllp{4.5cm}}     
\hline
ST & COR1-A & COR2-A & Flare & LASCO & Comments on the source \\
& time (UT) & time (UT) &   & CME  & \\
\hline
1 & 02:20 (55) & 02:54 & FL1 & & Small eruption to the south--east of AR 11123\\
2 & 05:00 (25) & 05:39 & FL2 & & Northern portion of filament F3 erupts from AR 11121\textsuperscript{*}\\
3 & 07:25 (45) & 07:54 & FL4 & SO1 & Eruption of a portion of F2 to the south of AR 11123\textsuperscript{*}\\
4 & 10:05 (10) & 10:24 & FL5 & & Jet from the north--east of AR 11123\\
5 & 10:25 (30) & 10:39 & FL6 & SO2 & Loops opening to the south--east of AR 11123\\
6 & 13:10 (40) & 13:24 & FL7 & SO3 & Eruption of F2 reformed to the south of AR 11123\textsuperscript{*}\\
7 & 13:50 (18) & 14:24 &  & & Jet from northern AR 11123\\
8 & 16:15 (22) & 16:39 & FL8 & SO4 & Eruption of filament F1\\
9 & 16:35 (45) & 16:54 & FL8 & SO4 & Eruption of southern portion of quiescent F3\textsuperscript{*}\\
10 & 19:35 (42) & 19:54 & FL9 & SO6 & Eruption of F2 reformed to the south of AR 11123\textsuperscript{*}\\
11 & 20:40 (15) & 21:08 & & & Faint jet to the east of AR 11123\textsuperscript{*}\\
12 & 21:45 (8) & 22:08 &  & SO7 & Faint jet from AR 11121\\
\hline
\multicolumn{6}{l}{\textsuperscript{*}\footnotesize{Ejections associated to filamentary material observed in 304\AA~images of SDO/AIA}}\\
\multicolumn{6}{l}{\footnotesize{and STEREO/SECCHI EUVI.}}\\
\end{tabular}
\end{table}

\begin{figure} 
\centerline{\includegraphics[width=1.\textwidth]{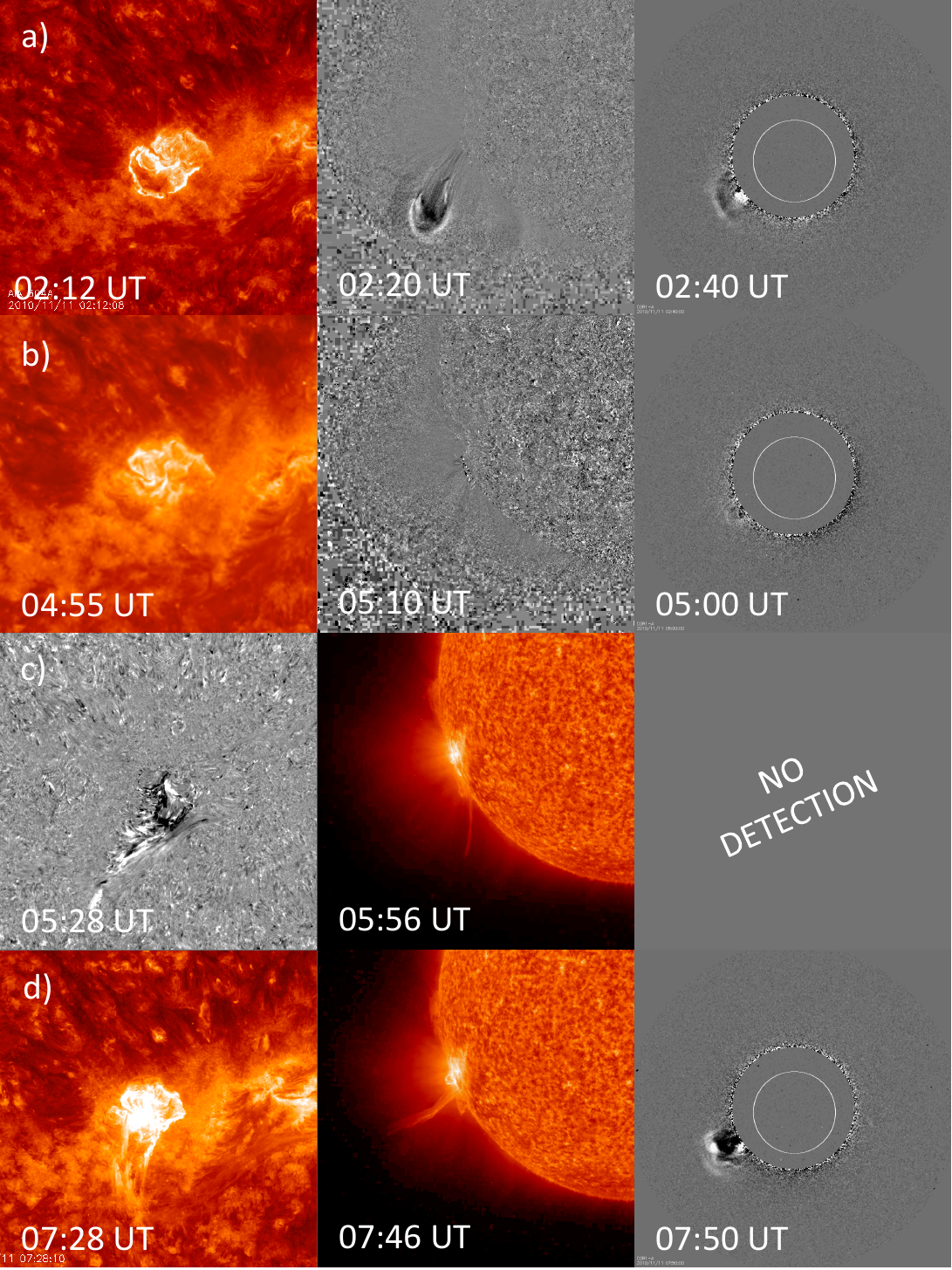}}
\caption{Eruptive events from the AR cluster. Rows (a) and (b) represent eruptive events ST1 and ST2 of Table~\ref{tbl:SECCHICMEs}. Row (c) refers to the eruptive activity associated to FL3, for which no white-light counterpart could be identified. Rows (d) to (m) represent eruptive events ST3 to ST12 of Table~\ref{tbl:SECCHICMEs}.  Left column: Earth's view of each event's source region, as registered in AIA 304\AA~(red-scale) or 193\AA~running-difference (grey-scale) images. Middle column: EUVI-A's quadrature view of the AR cluster with 304\AA~being red-scaled and 195\AA~running-difference grey-scaled. Right column: quadrature view of the CMEs from the COR1-A coronagraph (running-difference images).} 
\label{fig:collage1}
\end{figure}

\setcounter{figure}{3}
\begin{figure} 
\centerline{\includegraphics[width=1.\textwidth]{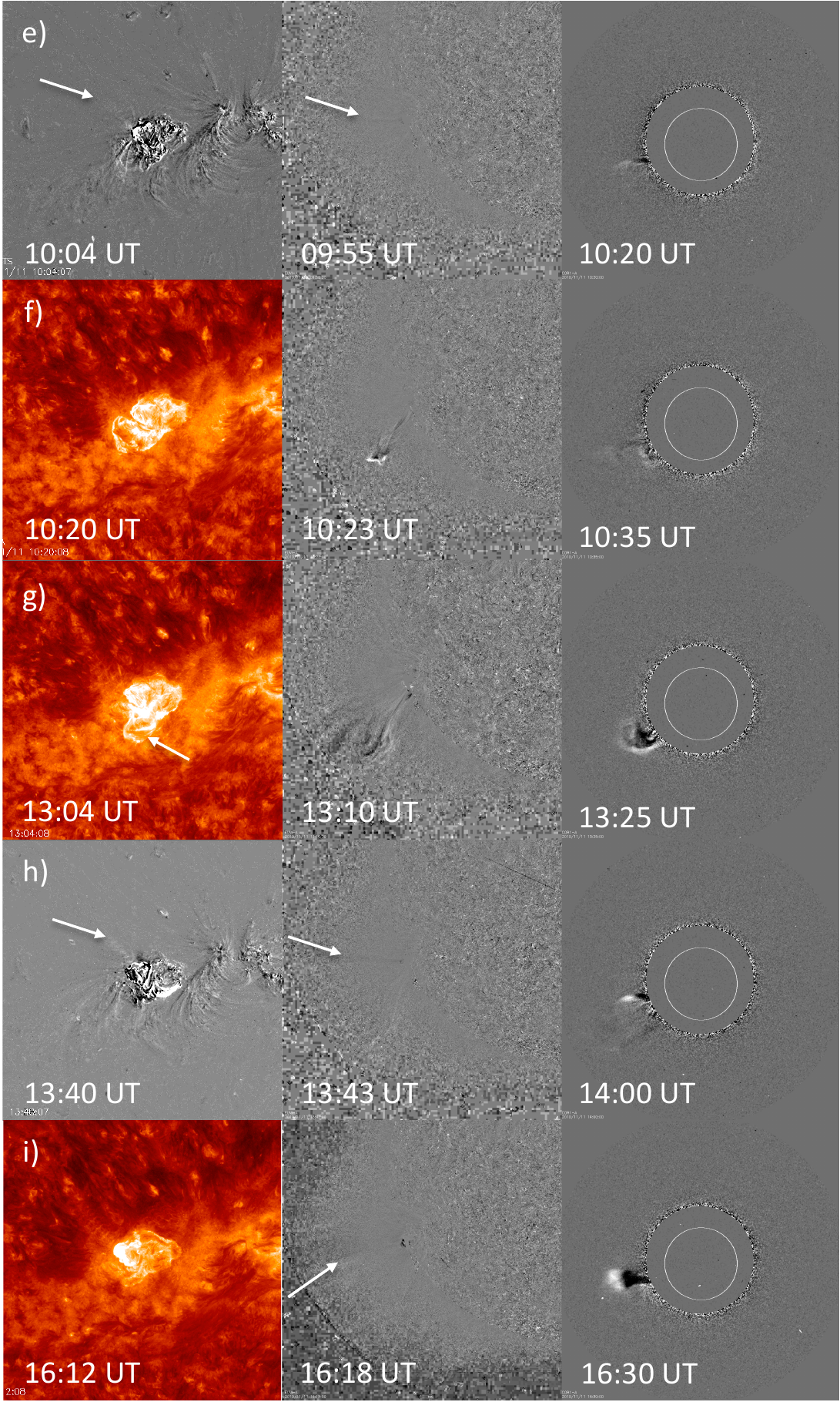}}
\caption{Continued.} 
\label{fig:collage2}
\end{figure}

\setcounter{figure}{3}
\begin{figure}
\centerline{\includegraphics[width=1.\textwidth]{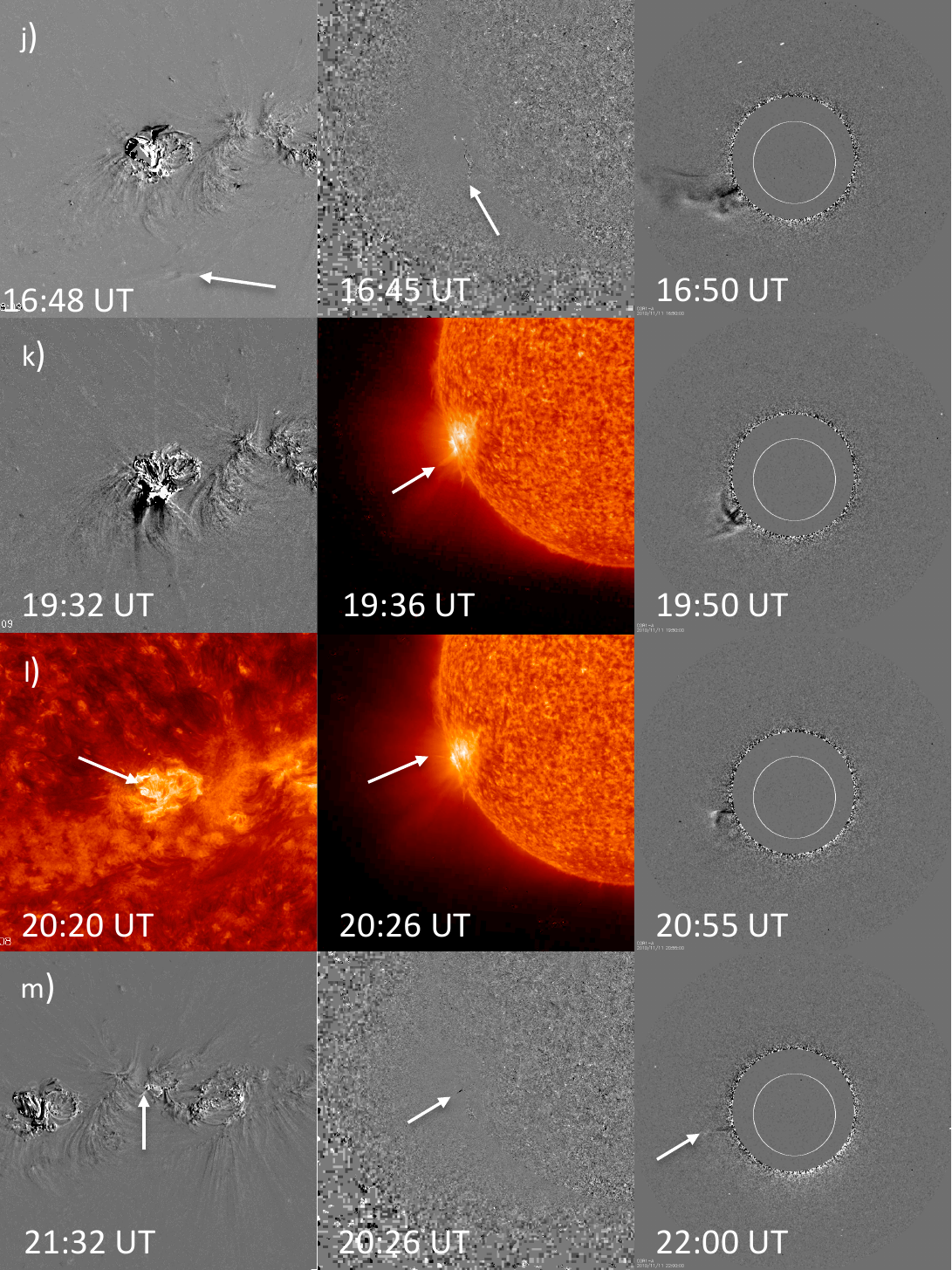}}
\caption{Continued.} 
\label{fig:collage3}
\end{figure}

Figure~\ref{fig:collage1} illustrates the eruptive events originating in the AR cluster. 
All STEREO/SECCHI CMEs were detected by both spacecraft but one, the very faint CME ST2 of Table ~\ref{tbl:SECCHICMEs}, only seen by the coronagraphs on the ST-A spacecraft (Figure~\ref{fig:collage1} row b). Although all of the CMEs seen by COR1 were later detected in the field of view of COR2 as well, some of them (ST1, ST2, ST5, ST11 and ST12) diffused in the background corona at relatively low heights, \ie below 5 \Rsun. Out of the 12 events listed in the table, four of them were narrow jets, less than 20$\degree$ wide. In some cases (events ST1, ST3, ST4, ST6, and ST10), it was not possible to discern whether the white-light features involved more than one ejected entity, leaving open the possibility for cases of two superimposed CMEs and cases of successive CMEs instead of trailing material. Because of these uncertainties, in Table~\ref{tbl:SECCHICMEs} we have considered these as single events, though it must be noted that the number of events on this day could be higher. The onset of all CMEs could be identified in the STEREO/SECCHI EUVI images, either in the 195\AA~or the 304\AA~wavelengths, or in both.  Once the low coronal signatures were identified from the STEREO perspective, SDO/AIA images were further inspected, constrained by location and timing considerations. This iterative process allowed a more precise determination of the CME sources in a synergistic use of multi-spacecraft data. During the onset of several of the eruptions (ST1, ST5, ST7, and ST10), large scale loops interconnecting ARs 11121 and 11123 were seen to open. Filamentary material was evident in 304\AA~images of SDO/AIA and STEREO/SECCHI EUVI for half of the events, which have been indicated by an asterisc in the last column of Table~\ref{tbl:SECCHICMEs}. 

Probably, the most remarkable finding after the data cross-matching is that every event reported by GOES as an X-ray flare (see Figure~\ref{fig:GOES} and the corresponding explanation in Section~\ref{GOES-description}) could be directly associated to an ejection. The weakest reported flares (see \eg flares FL5 and FL6  of Table~\ref{tbl:flares}, which after the initial examination of SDO/AIA data were not found to involve eruptions, had an ejective coronal counterpart, either in the form of a jet (CME ST4 of Table~\ref{tbl:SECCHICMEs}) or in the form of a bubble-like CME (CME ST5 of the same table). Although there was no white-light counterpart for flare FL3 of Table~\ref{tbl:flares} observed in the STEREO/SECCHI coronagraphs, the ejection of filamentary material is conspicuous in SDO/AIA and in STEREO/SECCHI EUVI images (see Figure~\ref{fig:collage1} row c). The matching of Tables~\ref{tbl:LASCOCMEs} and ~\ref{tbl:SECCHICMEs} yields two discords: CME SO4 of Table~\ref{tbl:LASCOCMEs} appears twice, while CME SO5 in the same table does not have a counterpart in Table~\ref{tbl:SECCHICMEs}. The first discrepancy is explained by the fact that STEREO/SECCHI CMEs ST8 and ST9 occur very close in time and are seen as one single event from the Earth's perspective. The second disparity  occurs because CME SO5 of Table~\ref{tbl:LASCOCMEs}, only detected by SEEDS, is most likely trailing material from STEREO/SECCHI CMEs ST8 and ST9. Also worth to note is the relatively narrow width of CME ST8, seen later as a partial halo CME from Earth's perspective.

\section{Discussion and Conclusions}\label{s:Discussion} 

The fast emergence of AR 11123 within the remnants of AR11121 close to CM and the series of flares taking place on 11 November 2010 pose this cluster as an important candidate for producing Earth-directed events. Its privileged location at CM makes it a perfect subject of study, though the CMEs arising from it were poorly observed from the Earth's perspective of SOHO/LASCO. Contrary to the expected, these CMEs appeared neither as full/partial halo nor as impressive events from Earth's perspective, except for one faint, ragged, partial halo CME. 

The difficult task of associating the eruptive activity observed on the cluster with the CMEs detected by SOHO/LASCO from Earth's perspective triggered a deeper study involving data from the twin STEREO spacecraft. When observing the cluster from a quadrature perspective, the STEREOs revealed almost twice as many CMEs, being some of them even massive and structured. This result is larger than the 32\% of frontside CMEs that cannot be recognized by SOHO found by \inlinecite{Wang2011}, and the findings by \inlinecite{Schwenn-etal2005}, who could not associate an observable halo CME to nearly 20\% of geoeffective interplanetary CMEs (ICMEs). On the other hand, after the analysis of 20 ICMEs detected in 2009 in the near-Earth solar wind by \inlinecite{Kilpua-etal2014}, only 6 events could be observed in SOHO/LASCO images, out of which only one was wider than 120$\degree$. After considering STEREO data, they were able to find the coronal counterparts of 16 ICMEs. Some of the associated CMEs were of the narrow type (less than 20$\degree$ wide), pointing out the importance of the narrow events identified in this study, given that they have the potential to arrive at Earth and exhibit clear \insitu~signatures. 

The STEREO/SECCHI EUVI instruments allowed the identification of limb low-coronal activity associated to the eruptions. Coronal events otherwise unseen from Earth's perspective also did not leave obvious imprints on SDO/AIA disk images of the low corona, posing these events as ``stealth''  CMEs (\opencite{Robbrecht-etal2009}).  However, all Earth-directed coronal events and their low coronal sources detected in STEREO/SECCHI EUVI could be associated subsequently, in the worst cases, to almost imperceptible activity on the best SDO/AIA images of the AR cluster at CM (see remarks on stealth CMEs in \opencite{Howard-Harrison2013}). \inlinecite{Kilpua-etal2014} remark that out of the 16 CMEs related with their set of 20 ICMEs, 10 were stealth, \ie with no obvious EUV on-disk activity. For the date of our case study, SDO was already operative, and the iteration between SDO/AIA and STEREO/SECCHI EUVI data was crucial to identify the precise location and time of each inconspicuous eruption.

Furthermore, the ejections found in STEREO/SECCHI quadrature observations indicate a direct flare-CME relationship if strictly speaking of the events reported by GOES as X-ray flares. Although these two phenomena are known to be related, they are not considered to be the cause of each other (\opencite{Webb2012}). It is well known that CMEs can occur without the company of flares, even when traveling at high speeds (\opencite{Song-etal2013}). On the other hand, several authors have found a majority of flares not associated to CMEs (\eg \opencite{Nitta-etal2014}). 
The most important findings can be summarized as follows:
\begin{enumerate}
\item[i)] Earth-directed CMEs did not always show up as halo or partial-halo CMEs in coronagraphs located in the Sun-Earth line.

\item[ii)] Quadrature observations revealed twice as many Earth-directed ejective events.

\item[iii)] Disk activity in the low corona did not suggest as many eruptions as were observed from the STEREO spacecraft.

\item[iv)] All events classified by GOES as X-ray flares on the studied date had an ejective coronal counterpart in quadrature observations.
\end{enumerate}

Although following the events on this day there was no geomagnetic storm reported, the implications for space weather forecasting are still straightforward; such an important number of missing alarms due to a lack of quadrature observations can be critical. Our results emphasize the need for the continuity of L4 or L5 space-weather dedicated missions, as also stated by \eg \cite{Lugaz-etal2010} and \cite{Gopalswamy-etal2011}, not only for the technology on which humankind depends but also for upcoming tripulated space missions of long duration.

%

%

%

%
\begin{acks}
HC and CHM acknowledge financial support from the Argentinean grants PICT 2012-973 (ANPCyT) and PIP 2012-100413 (CONICET). HC and CHM are members of the Carrera del Investigador Cient\'ifico (CONICET). AMC acknowledges funding from the Comisi\'on Nacional de Energ\'ia At\'omica (CNEA). 
The authors are thankful to Germ\'an Cristiani for his help with GOES data and to the reviewer for valuable comments and suggestions. The SOHO/LASCO data are produced by an international consortium of the NRL (USA), MPI f\"ur Aeronomie (Germany), Laboratoire d'Astronomie (France), and the University of Birmingham (UK). SOHO is a project of international cooperation between ESA and NASA. The STEREO/SECCHI project is an international consortium of the NRL, LMSAL and NASA/GSFC (USA), RAL and Univ. Bham (UK), MPS (Germany), CSL (Belgium), IOTA and IAS (France). SDO/AIA and SDO/HMI data are courtesy of the NASA/SDO and the AIA and HMI Science Teams.
This paper uses data from the SOHO/LASCO CME catalog generated and maintained at the CDAW Data Center by NASA and the CUA in cooperation with NRL, from the CACTus CME catalog generated and maintained by the SIDC at the ROB, the SEEDS project supported by the NASA/LWS and AISRP programs, and the LASCO ARTEMIS Catalog built by the Laboratoire d'Astrophysique de Marseille.

\end{acks}

%
%
\bibliographystyle{spr-mp-sola}
\bibliography{Cremades-etal}

\end{article} 
\end{document}